\RequirePackage{luatex85}
\documentclass[9pt,a4paper,twocolumn]{article}

\usepackage[utf8x]{inputenc}
\usepackage[english]{babel}
\usepackage[margin=1.5cm]{geometry}
\usepackage{times}
\usepackage{amsmath}
\usepackage{amsthm}
\usepackage{amssymb}
\usepackage{bm}
\usepackage{latexsym}
\usepackage{dsfont}
\usepackage{enumitem}

\usepackage{pgfplots} 
\definecolor{myblue}{rgb}{0.0,0.40,0.80}
\definecolor{myred}{rgb}{0.86,0,0.17}
\definecolor{mygray}{rgb}{0.3,0.3,0.3}

\usepackage{hyperref}
\hypersetup{
    colorlinks=true,
    linkcolor=blue, 
    citecolor=mygray, 
    urlcolor=blue} 

\usepackage{titlesec}
\titleformat*{\subsection}{\Large}

\newcommand{\T}[1]{\texttt{#1}}
\newcommand{\msun}{\mbox{$\mbox{M}_{\odot}$}}

\newcommand{\sub}[1]{\scriptsize\raisebox{-2pt}{$#1$}} 
\newcommand{\subt}[1]{\scriptsize\raisebox{-2pt}{#1}}  
\newcommand{\subD}{\tiny\raisebox{-2pt}{D}}

\title{
	\rule{\linewidth}{2pt} \\[0.5cm]
	\textsc{ARWV Code User Manual}
	\rule{\linewidth}{2pt}
}
\author{P. Chassonnery \thanks{E-mail: pauline.chassonnery@ens-cachan.fr}~, R. Capuzzo-Dolcetta  \thanks{E-mail: roberto.capuzzodolcetta@uniroma1.it}~ and S. Mikkola  \thanks{E-mail: mikkola@utu.fi}\\
ENS Paris-Saclay, France; Dep. of Physics, Sapienza, Univ. of Rome, Italy; Univ. of Turku, Finland
}

\date{October 2019}

\begin{document}
\maketitle

\section*{Introduction}

The \textsc{Fortran} file \T{ARWV\_v1.7.f} contains an updated and modified version of the algorithmic regularization code originally written by Seppo Mikkola. It is an open source program intended to integrate a few-body problem ($\T{N} \lesssim 500$) with high precision via the use of `chained variables' for the coordinates and velocities of the stellar bodies.  The set of equations of motions is that corresponding to the pair gravitational interaction (see \cite{RCDbook} for reference). In addition to its high computational accuracy, this program includes a number of interesting features among which Post-Newtonian (PN) terms in the computation of the acceleration (see \cite{Memmesheimer2004,Mikkola2008}). 

This program has been extended by other people. For example, an inclusion of the treatment of external regular potential and dynamical friction is due to M. Arca-Sedda \cite{ASCD19}.

Later, P. Chassonnery and R. Capuzzo-Dolcetta, in particular, have added a relativistic recoil velocity to the merger routine, and modified and improved the treatment of external, analytic, potentials and dynamical friction. 

The zipped package containing all the necessary files is downloadable
here:
\url{https://sites.google.com/uniroma1.it/}\\
\url{astrogroup/hpc-html}


Everyone is free to use it, upon proper citation of Mikkola's original papers and of this user manual.

\medskip

The \textsc{Fortran} file should be compiled with the command:
\begin{verbatim}
	> gfortran -o arwv.exe ARWV_v1.7.f
\end{verbatim}

To work, \T{arwv.exe} needs three files:
\begin{itemize}[leftmargin=3em]
	\item \T{ARCCOM2e2.CH}, which lists the \textit{common} variables and may not be modified except in a special case, as mentioned below;
	\item \T{PARAMETERS.TXT}, a set of working parameters;
	\item the data file \T{INPUT.TXT}, which contains the initial conditions (mass, position and velocity) for the set of \T{N} bodies.
\end{itemize}

At a chosen frequency, the code produces an output file called \T{100000n.dat}, where $n$ is an integer incrementing from $0$ (thus, you will have \T{1000000.dat}, \T{1000001.dat}, etc.). These files are formatted as the original input file, with the mass, coordinates and velocity components of all the stellar bodies belonging to the system. The integration time corresponding to each data snapshot is written at the end of the file.

\begin{table*}[ht]
\begin{center}
\begin{tabular}{ccp{9cm}}
	Parameter & Type and range & Meaning \\ \hline
	\T{N}         & $1 < \T{N} \leqslant \T{NMX}$ & Number
 of stellar bodies in the system \\
	\T{Nbh}       & $0 \leqslant \T{Nbh} \leqslant \T{N}$ & Number of black-holes in the system \\
	\T{TypeOfUnits} & $0/1$ & Indicates if the input data are given in \textit{autonomous} ($0$) or astrophysical units ($1$) \\
	\T{TMAX}      & real $> 0.0$ & Time over which the integration is conducted \\
	\T{DeltaT}  & real $> 0.0$ & Output interval ($\T{DeltaT} = 0.2$ recommanded) \\
	\T{cmet}      & real(3) $\geqslant 0.0$ & Indicates the regularization method to use, must be different from $(0,0,0)$ ($\T{cmet} = (1,0,0)$ or $(1,10^{-16},0)$ recommanded). \\
	\T{Clight}    & real $\geqslant 0.0$ & Value of the speed of light in the autonomous units (relevant only if \T{TypeOfUnits} is set to $0$) \\
	\T{Ixc}       & $0/1/2$ & Indicates if exact output times are required ($\T{Ixc} = 1$ recommanded) \\
    \T{ComputeOrbitalElements}  & yes/no & If yes, the code compute the keplerian orbital elements of the trajectory of each body with respect to the first body \\
	\T{tolerance} & real $\geqslant 10^{-14}$ & Tolerance in the accuracy of the integration ($\T{tolerance} = 10^{-12}$ recommanded) \\
	\T{soft}      & real $\geqslant 0.0$ & Softening of the potential ($\T{soft} = 0$ recommanded) \\
	\T{DehnenYes}, \T{DehnenDF} & yes/no ; yes/no & Activate Dehnen external potential and dynamical friction \\
	$\gamma_{\subD}$, $\T{r}_{\subt{D}}$, $\T{M}_{\subt{D}}$ & reals $\geqslant 0.0$ & Slope, scale radius and total mass of the Dehnen profile (ignored for \T{DehnenYes} = no) \\
	\T{PlummerYes}, \T{PlummerDF} & yes/no ; yes/no & Activate Plummer external potential and dynamical friction \\
	$\T{r}_{\subt{P}}$, $\T{M}_{\subt{P}}$ & reals $\geqslant 0.0$ & Scale radius and total mass of the Plummer profile (ignored for \T{PlummerYes} = no) \\
	\T{VirialYes}, \T{Q} & yes/no ; real $> 0.0$ & If yes the system is rescaled to the indicated virial ratio \T{Q} (rescaling to $\T{Q} = 1$ recommanded)
\end{tabular}
\end{center}
\caption{Overview of the user-defined parameters required by the code, as they are organized in the file \T{PARAMETERS.TXT}. For details on the available choices see the corresponding section.}
\label{Table_param}
\end{table*}

\section*{Input data}

The file \T{INPUT.TXT} is an ASCII file which must be provided by the user. It is composed of \T{N} rows, each giving the mass $m_i$, the cartesian coordinates $x_i,y_i,z_i$, and the cartesian velocity components $v_x^i, v_y^i, v_z^i$, of every body ($i=1, 2, \dots, N$) in the system. The order in which the data are given is arbitrary, except that the first \T{Nbh} rows refer to \textit{black-holes}. 

At the beginning of the code execution, the coordinates and velocities of the bodies are rescaled so that the center of mass of the system is in the origin of the coordinate frame and has a zero velocity. It is worth noticing that the external gravitational field is always centered in the center of mass of the \T{N}-body system. \\

The input data units can be chosen by the user. One possibility is by setting in the last line of \T{PARAMETERS.TXT} (see Table \ref{Table_param}) the parameter \T{TypeOfUnits} to $0$, so that the units of the entries in the file \T{INPUT.TXT} are ``free'' as long as the universal constant of gravitation, $G$, is set to $1$. We call these units \textit{autonomous}. The major constraint is that the parameters \T{TMAX}, \T{DeltaT}, \T{Clight} and, if relevant, $\T{r}_{\subt{D}}$, $\T{M}_{\subt{D}}$, $\T{r}_{\subt{P}}$ and $\T{M}_{\subt{P}}$ (see Table \ref{Table_param}) must be given in the same user-defined units.

If the user chooses \textit{astrophysical} units, the parameter \T{TypeOfUnits} must be set to $1$, and the various input data and physical parameters have to be given in solar masses (\msun), parsecs (pc) and km/s. The value of \T{Clight} is consequently $\sim 299,792$ and any value written for it inside \T{PARAMETERS.TXT} is not considered (the user must still write a dummy argument, to preserve the file format). To avoid both too large and too small numbers (with associated round-up error issues), the code will automatically converts all input data and parameters to the standard \T{N}-body units (for more details, see \cite{HeggieHut} p.$7$), defined as follow~:
\begin{itemize}
	\item the length unit is the initial size of the stellar system, i.e. $U_{\subt{l}} = \max\limits_{1\leqslant i \leqslant \T{N}}\sqrt{x_i^2 + y_i^2 + z_i^2}$;
	\item the mass unit is the total mass (\T{N} bodies + diffuse matter) inside the sphere of radius $U_{\subt{l}}$, that is 
	\begin{equation*}
		U_{\subt{m}} = \sum\limits_{i=1}^N m_i +
		\int_{V} \rho(\mathbf{r})\,d^3\mathbf{r},
	\end{equation*}
	where $\rho(\mathbf{r})$ is the density distribution generating, via solution of the Poisson's equation, the external potential;
	\item to ensure $G = 1. U_{\subt{l}}^3 U_{\subt{m}}^{-1} U_{\subt{t}}^{-2}$, the time scale must be $U_{\subt{t}} = \sqrt{U_{\subt{l}}^3/(G U_{\subt{m}})}$;
	\item accordingly, the unit of velocity is $\displaystyle U_{\subt{v}} = \frac{U_{\subt{l}} }{ U_{\subt{t}} } = \sqrt{\frac{ G U_{\subt{m}} }{ U_{\subt{l}} }}$.
\end{itemize} 

Note that \T{N}-body units are a peculiar case of autonomous units.
	
The output data will \textit{always} be provided in the same units than the input data.

\section*{Number and type of gravitating bodies}

\subsection*{Number of stellar bodies}

The number \T{N} of stellar bodies must be $\geqslant 2$. It is also capped at $\T{NMX} = 400$ (see file ARCCOM2e2.CH).

Although this is not recommended, the upper limit \T{NMX} can be changed. To do so, one has first to modify the variable \T{NMX} to the wanted value in the file \T{ARCCOM2e2.ch} and the routine \T{NotAtAllNeeded} of the main file. 

Then, the variable with same name \T{NMX} present in the routines \T{ARC}, \T{Iterate2ExactTime} and \T{DIFSYAB} of the main code must be modified accordingly : if the common variable \T{NMX} is (for example) multiplied by $2$, then the user must also multiply the homonimous variable by $2$ in the three routines indicated above.

\subsection*{Number of black-holes}

The code discriminates the \T{N} stellar bodies into two populations~: the ``classic'' stars and the ``black-holes'', these latter being the \T{Nbh} first bodies entered into the file \T{INPUT.TXT}. The motion of a classic star is governed by the standard Newtonian equations, while for black-holes a PN approximation up to $2.5$ order is adopted (see \cite{Memmesheimer2004} for details of the equations and \cite{Mikkola2008} for their implementation).

The number \T{Nbh} of black-holes may range from $0$ (no PN effect at all) to \T{N} (all stellar bodies are black-holes). However, \T{Nbh} greater than \T{N} will be accepted by the code and interpreted as $\T{Nbh} = \T{N}$.

To compute the PN terms, the code needs the value of the speed of light in the units used for the simulation. If the input data are entered in physical units ($\T{TypeOfUnits} = 1$), then this value is computed simultaneously at the conversion into \T{N}-body units, with:
$$\T{Clight} = \frac{2.99792\times 10^5 \mbox{ km/s}}{U_{\subt{v}}}.$$

If input data are given in autonomous units ($\T{TypeOfUnits} = 0$), the user must indicate in the file \T{PARAMETERS.TXT} the correct value of \T{Clight} in these units. It has to be noted that setting $\T{Clight} = 0$ will switch off all PN terms and effects (including the merger process, see related section), even if $\T{Nbh} \neq 0$. Doing things this way can prove a little tricky for the user~: it is more straight-forward to set $\T{Nbh} = 0$. Note, also, that setting \T{Clight} at a very large value makes the role of PN terms negligible at a useless comptutational cost.

When removing the PN part by setting $\T{Nbh} = 0$, the integration proceeds much faster. On the other hand, the simulation as a whole can get utterly slow if there is any close two-body encounter, because the code cannot proceed to a merger; then, to ensure sufficient accuracy in the modelization of the binary motion, the time-steps reduce enormously and the simulation process get stuck.

\section*{Time-related parameters}

\subsection*{Total time of integration}

The input variable \T{TMAX} represents the total time over which the evolution of the system will be simulated. If $\T{TypeOfUnits} = 0$, its value is assumed to be in the autonomous time unit of the system. If $\T{TypeOfUnits} = 1$, it is considered to be given in million years ($10^6$ yrs).

\subsection*{Output interval} 

While computing the evolution of the system, the code uses an adaptative time-step chosen so as to ensure enough accuracy in the treatment of tight binaries (see the section about time-transformation). This time-step can be very small, so that writing down the state of the system at each step would both slow down the computation and produce an enormous amount of output files which could overflow the memory storage.

To avoid this, the user may define the output time interval, hereafter called \T{DeltaT}. If $\T{TypeOfUnits} = 0$, its value is assumed to be in the autonomous time unit of the system. If $\T{TypeOfUnits} = 1$, it is considered to be given in million years, like \T{TMAX}. It is worth noticing that \T{DeltaT} is also used for initializing the computation of the adaptative time-step. 

The parameter \T{Ixc} allows the user to select the way the output is triggered. Indeed, since the optimal integration time-step is computed independently by the code, the actual integration time $t$ is \textit{not} a priori a multiple of \T{DeltaT}.

Let $t_0$ be the integration time at which the last output file has been produced, and $t_{\subt{next}} = t_0 + \T{DeltaT}$ the time for the, possible, next output.
If $\T{Ixc} = 0$, then a new output file is written as soon as the integration has gone far enough -- that is when $t>t_{\subt{next}}$. This is the most straightforward method but, depending on the system, the actual global time interval can be much longer than the one the user asked for.

If $\T{Ixc} = 2$ on the other hand, the code will respect the exact output times required by integrating to $t = t_{\subt{next}}$. This can be slow if many outputs are required, that is if $\T{DeltaT} \ll \T{TMAX}$. 

A compromise between these two methods is provided by the choice $\T{Ixc} = 1$. In such case the code tries to keep the time interval close to the wanted \T{DeltaT} by approximating the exact output time $t_{\subt{next}}$ with the Stumpff-Weiss method \cite{stu67}. 
This option is often the fastest and most suitable one. \\

It must be noted that the setting $\T{Ixc} = 0$, and to a lesser extent $\T{Ixc} = 1$, works well only when \T{DeltaT} is reasonably small. If \T{DeltaT} is large, one may prefer the use of the slower $\T{Ixc}=2$ choice. However, the notion of what ``small'' is depends on the specific case.

\subsection*{Time-transformation or regularization}

In few-body systems, close approaches happen frequently. Consequently, to ensure accuracy in numerical simulations, one must {\it regularize} the strong interaction by a proper computation of coordinates and velocities (which is done by the leapfrog algorithm). Several methods exist for that purpose, which can efficiently be used, not only for the purely Newtonian few-body problem but also for other problems with external perturbations, including velocity dependent ones (in our case, external gravitational field with dynamical friction, this latter depending on the object speed).
\par\noindent
The \T{ARWV} code proposes a combination of three of them~: 
\begin{itemize}
	\item the {\it logarithmic-Hamiltonian} (logH) method, proposed by Mikkola and Tanikawa (see \cite{Mikkola99a} and \cite{Mikkola99b}) as well as Preto and Tremaine \cite{PretoTremaine}, gives accurate trajectories for the two-body problem and also satisfactory results for perturbed two-body encounters; 
	\item the Time-Transformed-Leapfrog (Mikkola and Aarseth \cite{Mikkola2002}), which turned out to be close to the logH-method, but, in some respects, more general;
	\item the auxiliary velocity algorithm, presented by Hellstr\"{o}m and Mikkola in \cite{Mikkola2010}, which is not a real regularization method, but rather  a trick which allows to apply the ``normal'' leapfrog algorithm to a velocity-dependent acceleration.
\end{itemize}

The parameter \T{cmet} is a three-dimensional vector that determines the way these methods will be combined. $\T{cmet} = (1,0,0)$ is the logarithmic-Hamiltonian, $(0,1,0)$ is the Time-Transformed-Leapfrog and $(0,0,1)$ corresponds to no time-transformation.

Setting $\T{cmet} = (1,0,0)$ is usually the best choice, but a small value for \T{cmet}(2) may be advisable in case of star-star close encounter. If the system contains big and small bodies, one may use $\T{cmet} = (1,10^{-16},0)$ instead of $(1,0,0)$ : this causes smaller stepsizes when two very small bodies come close to each other. Please note that, in this case, the computational speed depends heavily on the formation of binaries !

\section*{Merging and recoil velocity}

The code includes a \textit{merging} procedure intended for the fusion of a black-hole binary or for the swallowing of a classic star by a black-hole (but not for collisions between two stars; this means that collisions are handled out only if at least one of the two objects is in the list of the \T{Nbh} black-holes).

The classic recipe for the merging of two bodies is that the remnant of the merger will be located at the center of mass of the previous pair (considered as the location where the collision actually happens, though the code stops the integration of the separate trajectories just before that point).

For the velocity of the remnant (or ``recoil'' velocity) however, there is no general consensus. For a purely Newtonian simulation, possible alternatives are giving to the remnant a null velocity or the velocity of the center of mass of the two progenitors. In the original version of \T{ARWV}, the latter choice has been done. \\

The code was later modified by Roberto Capuzzo-Dolcetta and Pauline Chassonnery to follow the prescription given by Lousto and Healy in \cite{LoustoHealy}. 

Let $m_{\sub{1}}$ and $m_{\sub{2}}$ be the masses of two merging black-holes. We use the convention $m_{\sub{1}} \leqslant m_{\sub{2}}$ and introduce the auxiliary variables:
\begin{align*} 
	&m = m_{\sub{1}} + m_{\sub{2}}, \hspace{1em} \eta = \frac{m_{\sub{1}} m_{\sub{2}}}{m^2}, \hspace{1em} 0< q = \frac{m_{\sub{1}}}{m_{\sub{2}}} \leqslant 1, \\
	&\delta_m = \frac{m_{\sub{1}} - m_{\sub{2}}}{m} \leqslant 0.
\end{align*}

Each stellar body is assumed to be auto-rotating, with a dimensionless spin $\bm{\alpha}_i$ such that $|\bm{\alpha}_i| \leqslant 1$. Two spin-related variables are~: 
\begin{align*}
	\bm{S} &= \frac{ \bm{\alpha}_{\sub{1}} m_1^2 + \bm{\alpha}_{\sub{2}} m_2^2 }{m^2} = \frac{ \bm{\alpha}_{\sub{2}} + q^2 \bm{\alpha}_{\sub{1}}}{(1 + q)^2}, \\
	\bm{\Delta} &= \frac{ \bm{\alpha}_{\sub{2}} m_{\sub{2}} - \bm{\alpha}_{\sub{1}} m_{\sub{1}} }{m} = \frac{ \bm{\alpha}_{\sub{2}} - q \bm{\alpha}_{\sub{1}}}{1 + q}.
\end{align*}
$ $

The dimensionless spin of each of the \T{N} bodies can be generated randomly at the beginning of the simulation, or read from a separated file. The latter is best for reproductibility purpose, but need the user to provide a file named \T{SPIN.TXT}, formatted as \T{N} rows containing each the three coordinates of the spin of one body. The user must also uncomment, in the main routine of the code, the part from line 157 to 182 included, i.e. the section "SET INITIAL SPINS". \\

The recoil velocity is finally given by:
$$\textbf{v}_{\subt{recoil}} = v_{\subt{m}} \textbf{e}_{\sub{1}} + v_{\sub{\perp}} \left( \cos \xi\textbf{e}_{\sub{1}} + \sin \xi\textbf{e}_{\sub{2}} \right),$$
where $\textbf{e}_{\sub{1}}$ is the unit vector pointing from $m_{\sub{1}}$ to $m_{\sub{2}}$ and $\textbf{e}_{\sub{2}}$ an orthogonal unit vector in the two-body orbital plane, such that the basis formed by $\textbf{e}_{\sub{1}}$, $\textbf{e}_{\sub{2}}$ and the orbital angular momentum of the pair is direct.

The values of $\xi$, $v_{\subt{m}}$ and $v_{\sub{\perp}}$ are given by the following fitting formulas to numerical data:
\begin{align*}
	&\xi = a + b S_{||} + c \delta_m \Delta_{||}, \\
	&v_{\subt{m}} = \eta^2 \delta_m \left(A + B \delta_m^2 + C \delta_m^4 \right), \\
	&v_{\sub{\perp}} = H \eta^2 \left( \Delta_{||} + H_{\subt{2a}} S_{||} \delta_m + H_{\subt{2b}} \Delta_{||} S_{||} \right. \\
	 &\hspace{2em} + H_{\subt{3a}} \Delta_{||}^2 \delta_m + H_{\subt{3b}} S_{||}^2 \delta_m + H_{\subt{3c}} \Delta_{||} S_{||}^2 \\
	 &\hspace{2em} + H_{\subt{3d}} \Delta_{||}^3 + H_{\subt{3e}} \Delta_{||} \delta_m^2 + H_{\subt{4a}} \Delta_{||}^2 S_{||} \delta_m \\
	 &\hspace{2em} + H_{\subt{4b}} S_{||}^3 \delta_m + H_{\subt{4c}} S_{||} \delta_m^3 + H_{\subt{4d}} \Delta_{||} S_{||} \delta_m^2 \\
	 &\hspace{2em} \left. + H_{\subt{4e}} \Delta_{||} S_{||}^3 + H_{\subt{4f}} \Delta_{||}^3 S_{||} \right).
\end{align*}
where the subscript $||$ denotes the projection of vectors along the direction of the orbital angular momentum of the pair. The numerical values of the different constants above are given in \cite{LoustoHealy}. Note that $A$, $B$, $C$ and $H$ are given in km/s and must be converted into the units used for the simulation. This is done automatically by the code, no matter the value of \T{TypeOfUnits} (unless $\T{Nbh} = 0$ and/or $\T{Clight} = 0$, in which cases PN terms are switched off and mergers are not permitted).

\section*{External gravitational fields}

This code allows the user to add to the simulated stellar system an external regular and spherical distribution of matter, accounting for its gravitation field and dynamical friction over the \T{N} bodies.

The code allows two possible density profiles for the external field : a {\it Dehnen} profile \cite{Dehnen93} and/or a {\it Plummer} profile \cite{Plummer}. The user may activate either of these by setting to ``yes'' the corresponding option \T{DehnenYes} and/or \T{PlummerYes} (the two options can be activated at once, to get a superposition of the two density profiles).

The matter distribution is defined by its total mass (parameter $\T{M}_{\subt{D}}$ or $\T{M}_{\subt{P}}$) and its scale radius (parameter $\T{r}_{\subt{D}}$ or $\T{r}_{\subt{P}}$). These two values may be given either in $\msun$ and parsec (option $\T{TypeOfUnits} = 1$) or in the same autonomous units than the other input data (option $\T{TypeOfUnits} = 0$). For the Dehnen profile, the user also needs to define a central slope (parameter $\gamma_{\subD}$), whose value must be $0\leqslant \gamma_{\subD}< 5/2$. \\

The spherical matter distribution acts over the \T{N} objects by mean of its gravitational field and dynamical friction braking.  

The dynamical friction (df) effect is taken into account via a deceleration term in the classical Chandrasekhar’s form in local approximation \cite{Chandrasekhar}: its magnitude depends upon a proper average of the velocity dispersion of the particles composing the regular matter distribution. For a Plummer profile, the average velocity dispersion is :
$$\sigma^2_{\subt{P}} = 3 \pi G M_{\subt{P}}/(32 r_{\subt{P}}).$$

For a Dehnen profile, the formula for the gravitational energy converges only if $\gamma_{\subD} < 5/2$, to give the following expression :
$$\sigma^2_{\subt{D}} = \frac{G M_{\subt{D}}}{2 r_{\subt{D}}} \frac{1}{5-2\gamma_{\subD}}.$$

Then, for the superposition of a Dehnen and a Plummer profile, the global mass-averaged velocity dispersion is:
$$\sigma^2 = \frac{M_{\subt{D}} \sigma^2_{\subt{D}} + M_{\subt{P}} \sigma^2_{\subt{P}}}{M_{\subt{D}} + M_{\subt{P}}}.$$

\section*{Initial virial ratio}

A physical system is said to be in virial equilibrium if the second derivative $\ddot{I}$ of its polar moment of inertia is null, at least on average. 

Let $K$ denote the kinetic energy of a system and $\Omega$ its potential energy. We can show that $\ddot{I}/2 = 2K +\Omega$ (see \cite{HeggieHut}, chapter $9$ for proof).
Calling ``virial ratio'' the quantity $Q = 2K/|\Omega|$, the condition $\langle\ddot{I}\rangle = 0$ is equivalent to $\langle Q \rangle = 1$. \\

When studying systems embedded in a regular mass distribution, the gravitational attraction exerted by the regular field is, generally, non negligible and deserves to be included in the computation of the total potential energy $\Omega$ of the system.

Let $Q_{\sub{0}}$ denote the virial ratio corresponding to the input data given by the user (file \T{INPUT.TXT}). If the parameter \T{VirialYes} (in the file \T{PARAMETERS.TXT}) is set to ``yes'', then the code  rescales the system's velocities so that the virial ratio gets the value \T{Q} assigned by the user in the \T{PARAMETERS.TXT} file. This is, simply, done by multiplying the velocities of all the stellar bodies by $\sqrt{\T{Q}/Q_{\sub{0}}}$. \\

\section*{Keplerian orbital elements}

An interesting case of \T{N}-body system are systems where small bodies (stars as well as black-holes) are moving around a massive rotating black-hole. Most of the small bodies can then be considered as orbiting around the central one in a binary relationship (like the planets around the Sun). In this case, the user may be more interested in the characteristics of these binary-orbits (semi-major axis, eccentricity, period...) than by the raw coordinates and velocities of each body.

The user may thus place the data of the massive black-hole in the first line of the file \T{INPUT.TXT} and activate the binary motion analysis by setting to ``yes'' the option \T{ComputeOrbitalElements} in the parameters file.

In PN theory, the characteristic variables of a binary motion, or orbital elements, depend on the spin of the two members of the binary. If we assume a large unbalance between the masses, the spin of the small body can be neglected and the code only require the spin of the massive black-hole. 

This (dimensionless) spin $\bm{\alpha}_{\sub{1}}$ has already been defined, since it is necessary to compute the recoil velocity (see related section). We recall here that this spin can be defined via a file \T{SPIN.TXT} provided by the user, or randomly generated by the code at its initialization. It must have a norm less than $1$ and, unless the spin of the other bodies which are constant, it will evolve along the computation.

At each output time, the code will compute the value of the orbital elements for all pairs $(m_{\sub{1}},m_{\sub{j}})_{1<j\leqslant N}$ and save them in separate files numbered from $71$ to $76$~:
\begin{itemize}
	\item file \T{71.dat} contains the semi-major axis $a$ of each pair at each output time,
	\item file \T{72.dat} contains the eccentricity, $e$, of each pair, 
	\item file \T{73.dat} contains the inclination $i$ (or vertical tilt of the orbital plane with respect to the reference plane,
	\item file \T{74.dat} contains the longitude of the ascending node $\Omega$ (or orientation of the orbital plane with respect to the reference frame's vernal point),
	\item file \T{75.dat} contains the argument of periapsis $\omega$ (or orientation of the ellipse in the orbital plane),
	\item file \T{76.dat} contains the time, the current spin $\bm{\alpha}_{\sub{1}}(t)$ of the first body, and the absolute variation of the norm of this spin (i.e. $\alpha_{\sub{1}}(t) - \alpha_{\sub{1}}(0)$).
\end{itemize}

\section*{Miscellanous}

\subsection*{Tolerance}
The parameter \T{tolerance} sets the integration accuracy in the routine \T{DIFSYAB}. A small value  is suggested, but not less than $10^{-14}$ (if \T{tolerance} is smaller, the code sets it at $10^{-14}$). Of course, the code is faster when requiring low accuracy, but it must be remembered that \T{ARWV} is designed to conduct highly accurate simulations.

\subsection*{Softening}
The parameter \T{soft} allows the user to {\it smooth} the straight $1/r$ Newtonian potential with one of the form $1/\sqrt{r^2 + \T{soft}^2}$. To have a purely Newtonian interaction this parameter must be set to $0$.

\section*{Acknowledgements}
We thank M. Arca-Sedda for providing us a version of the archain code with some modifications he made, that constituted a starting version form our further changes and additions which eventually lead to the version described in this user manual and that can be downloaded.

\end{document}